\documentclass[sigconf]{acmart}

\usepackage{mdframed}
\usepackage{csquotes}
\usepackage{dblfloatfix}
\usepackage{listings}
\usepackage{cuted}
\usepackage{caption}
\usepackage{graphicx}
\usepackage{subcaption}
\usepackage{xcolor}
\usepackage{float}
\usepackage{tabularx}
\usepackage{makecell}
\usepackage{balance}
\usepackage[inkscapelatex=false]{svg}
\usepackage{afterpage}
\usepackage{caption}
\usepackage{placeins}
\usepackage{float}

\AtBeginDocument{%
  }

\copyrightyear{2026}
\acmYear{2026}
\setcopyright{cc}
\setcctype{by}
\acmConference[ReCode '26]{1st Workshop on Code Translation, Transformation, and Modernization }{April 12--18, 2026}{Rio de Janeiro, Brazil}
\acmBooktitle{1st Workshop on Code Translation, Transformation, and Modernization (ReCode '26), April 12--18, 2026, Rio de Janeiro, Brazil}
\acmPrice{}
\acmDOI{10.1145/3786180.3788316}
\acmISBN{979-8-4007-2411-4/2026/04}

\begin{document}

\title{JMigBench: A Benchmark for Evaluating LLMs on Source Code Migration (Java 8 to Java 11)}

\author{Nishil Amin}
\affiliation{%
  \institution{University College London}
  \city{London}
  \country{United Kingdom}}
\email{nishil.enq@gmail.com}

\author{Zhiwei Fei}
\affiliation{%
  \institution{Nanjing University}
  \city{Nanjing}
  \country{China}}
\email{zhiweifei@smail.nju.edu.cn}

\author{Xiang Li}
\affiliation{%
  \institution{University College London}
  \city{London}
  \country{United Kingdom}}
\email{x.li.25@ucl.ac.uk}

\author{Justyna Petke}
\affiliation{%
  \institution{University College London}
  \city{London}
  \country{United Kingdom}}
\email{j.petke@ucl.ac.uk}

\author{He Ye}
\authornote{Corresponding author.}
\affiliation{%
  \institution{University College London}
  \city{London}
  \country{United Kingdom}}
\email{he.ye@ucl.ac.uk}

\renewcommand{\shortauthors}{Nishil Amin et al.}

\begin{abstract}
We build a benchmark to evaluate large language models (LLMs) for source code migration tasks, specifically upgrading functions from Java 8 to Java 11. We first collected a dataset of function pairs from open-source repositories, but limitations in data quality led us to construct a refined dataset covering eight categories of deprecated APIs. Using this dataset, the Mistral Codestral model was evaluated with CodeBLEU and keyword-based metrics to measure lexical and semantic similarity as well as migration correctness. Results show that the evaluated model (Mistral Codestral) can handle trivial one-to-one API substitutions with moderate success, achieving identical migrations in 11.11\% of the cases, but it struggles with more complex migrations such as CORBA or JAX-WS. These findings suggest Mistral Codestral can partially reduce developer effort by automating repetitive migration tasks but cannot yet replace humans within the scope of the JMigBench benchmark. The benchmark and analysis provide a foundation for future work on expanding datasets, refining prompting strategies, and improving migration performance across different LLMs.
\end{abstract}



\maketitle

\section{Introduction}
Modern software development requires continuous evolution of codebases to incorporate new features, address security issues, and ensure maintainability~\cite{rewardrepair, selfapr, ye2024iter, ye2025adverintent}. At the same time, programming languages progress through successive versions, introducing new features while deprecating or removing existing APIs. This evolution forces developers to migrate legacy code to remain compatible with new versions. Failure to perform timely migration can lead to technical debt, increased maintenance costs, and security vulnerabilities.

Source code migration is both essential and burdensome. Prior studies at Google report that “developers do not find the work rewarding” and that “automation is preferred for such migrations”~\cite{MigratingGoogle}. Automating this process could significantly reduce developer workload. Rule-based tools~\cite{VISSER2005831, CORDY20043} and machine learning approaches~\cite{huang2023distill} have been developed, but they either lack the flexibility to handle complex changes or require large, structured training datasets. With the advent of large language models (LLMs), researchers and practitioners are beginning to explore whether LLMs can automate migration tasks at scale~\cite{codemenv, migrationbench}.

However, progress is limited by the lack of suitable benchmarks. Existing datasets such as HumanEval~\cite{chen2021evaluatinglargelanguagemodels} and SWE-bench~\cite{2023evalllmnk} focus on code generation and bug fixing rather than language-version migration. CODEMENV~\cite{codemenv} considers migration but only in narrow scenarios. MigrationBench~\cite{migrationbench} evaluates repository-level migration from Java 8 to newer versions, but it does not provide ground-truth function pairs, making fine-grained evaluation of migration correctness difficult. This leaves a critical gap: there is no benchmark for evaluating whether LLMs can correctly migrate code between language versions at the function level.

While newer long-term support (LTS) versions of Java exist, such as Java 17 and Java 21, this paper focusses on migration from Java 8 to Java 11. Java 11 was the first LTS release following Java 8 and it therefore represents a meaningful evolution in the Java platform. A developer survey in 2024 highlighted that \enquote{two-thirds of respondents admitted to still running applications on Java 11 or earlier versions} \cite{javaversionuses}. This makes Java 8 to 11 a valuable benchmark to use when evaluating LLMs as a migration technique.

Constructing such a dataset is challenging, as deprecated APIs are inconsistently used across repositories, often appear in unrelated contexts, and require careful alignment of pre- and post-migration code.

We design this benchmark to evaluate LLMs on real-world migration tasks, focusing on the Mistral Codestral model. Using CodeBLEU and keyword-based metrics, we measure lexical and semantic similarity as well as migration correctness. Our findings show that Mistral Codestral can automate trivial one-to-one substitutions but struggles with more complex migrations where CORBA and JAX-WS APIs were involved.

Our contributions are as follows:
\begin{itemize}
    \item We build a benchmark of Java 8 to Java 11 function pairs migration across eight categories of deprecated APIs, which is publicly available at \url{https://github.com/NishilAmin1213/JMigBench}.
    \item A comparative analysis of LLM outputs against human-written Java 11 code, quantifying lexical and semantic similarity between them.
    \item Empirical insights showing that while LLMs can reduce developer effort perfectly on 11.11\% of the trivial cases, with a keyword removal success rate of 31.82\%, human intervention remains essential for complex migrations, such as the JAX-WS and CORBA APIs where the LLM failed to remove any meaningful keywords. 
\end{itemize}
The remainder of this paper is organised as follows. Section~\ref{sec:related_work} presents the background and related work, including current benchmarks on code migration and existing migration techniques. Section~\ref{sec:Jmig_bench} details the construction methods and the information about this dataset. Section~\ref{sec:exp_method} details the experimental methodology, including the research questions, evaluation metrics, and prompting strategies. Section~\ref{sec:results} reports the experimental results across two research questions. Section~\ref{sec:threats} discusses threats to validity. Finally, Section~\ref{sec:conclusion} concludes the paper and summarises the key findings.

\section{Related Work}\label{sec:related_work}

Existing benchmarks for code migration remain limited. We review and compare current benchmarks with our proposed one and provide an overview of the current state of migration techniques.

\subsection{Code Migration Benchmarks}

The current code migration benchmarks are limited. To the best of our knowledge, there are two datasets that focus on this direction.

\textbf{CODEMENV} is a benchmark which is \enquote{designed to assess LLMs' abilities in code migration scenarios}. It contains 922 test cases covering 3 scenarios, \enquote{identifying functions incompatible with specific versions}, \enquote{detecting changes in function definitions} and, \enquote{adapting code to target environments} \cite{codemenv}.

\textbf{MigrationBench} is the first large scale benchmark to evaluate code migration from Java 8 to newer LTS versions (Java 17 and Java 21). The project scraped over 5,000 repositories from GitHub to build its dataset. By working at a repository level, Liu et el. were able to measure success through criteria such as successful compilation, dependency resolution, and the execution of test suites. The paper provided a baseline using tools such as OpenRewrite enabling accurate reflection on the ability of LLMs \cite{migrationbench}.
\\
Both MigrationBench and our work share the goal of evaluating LLMs on real world migration tasks, but they have a range of differences. This work focuses on migration to Java 11. Although it is an older version than Java 17 or 21, it remains a critical upgrade due to its status as the first LTS version after Java 8, and the fact that Java 11 has seen widespread adoption in the industry. As opposed to the repository level approach of MigrationBench, this project narrows the scope down to a functional level, producing a more fine-grained benchmark of the LLMs ability to upgrade code, and focusing less on the LLMs ability to process configuration files. Another key change is the availability of ground truth data. The dataset this project aims to develop will pair Java 8 code with corresponding Java 11 code post migration. This will a direct comparison of generated and human written code. MigrationBench in contrast, evaluates success through build outcomes and test execution.

\subsection{Existing Migration Techniques}
Prior to the widespread adoption of LLMs, a suite of techniques existed to automate the task of source code migration: 
\\
\textbf{Rule-based techniques} apply \enquote{one step} transformations to code represented by syntax trees \cite{VISSER2005831}. Exemplified in tools such as TXL, Spoon and OpenRewrite \cite{CORDY20043, SPOON2015, openrewrite}. These are efficient when working with simple and repetetive changes, but can fail when handling more complex semantic changes. 
\\
\textbf{Machine learning methods} such as CoDist train models to \enquote{capture the semantic and structural equivalence of code}. Whilst this solves the drawbacks of rule based methods, it requires a large structured dataset to train models effectively \cite{huang2023distill}. 
\\
\textbf{Hybrid approaches} combine both of the aforementioned methods. Malyala et al. showed it to be advantageous as they measured an 86\% improvement over standalone ML techniques after sandwiching an unsupervised ML model between two rule-based processors \cite{malaya23translation}. Although they were converting code from Java to Python, the paper provides valuable insight into how similar techniques could be applied to migration between language versions.

\section{JMig Bench: Benchmark for Java 8 to 11 Migration}\label{sec:Jmig_bench}
To fill the current gap in measuring LLMs' ability in code migration, we propose \textbf{JMig Bench}, a benchmark of function pairs migrated from Java 8 to Java 11, which focuses on evaluating models' fine-grained, function-level capabilities. In this section, we detail the dataset construction process and provide its static information.
\subsection{Dataset Collection}

We focus on the migration from Java 8 to Java 11, as both are Long-Term Support (LTS) versions widely adopted in the industry. To construct a representative dataset, we selected popular open-source Java repositories on GitHub with over 10,000 stars to ensure code quality and active maintenance. For each repository, our program examined release notes, open issues, and recent commits to identify those containing explicit evidence of a Java 8 to Java 11 migration (using targeted keyword matching). Each candidate repository was then manually verified to confirm the migration event.

From the verified repositories, we paired corresponding Java 8 and Java 11 branch URLs and extracted function-level changes using our own python code. We only retained functions with the same name but non-identical implementations and with more than 10 lines of code, ensuring sufficient semantic content for analysis. Furthermore, we ensured that the Java 8 versions contained deprecated or removed syntax in Java 11 through keyword matching.

\subsection{Quality Checking and Synthesizing}
After filtering repositories by the aforementioned conditions, Jenkins and Bazel were the two prominent projects that remained. The raw dataset consists of two sub-datasets of corresponding Java 8 and 11 functions, one for 131 functions with the same function signature, and another for 19 functions with a different function signature. Combined, the dataset contains 150 functions from GitHub where:
\begin{itemize}
    \item The Java 8 function contains syntax that was removed by Java 11 
    \item A Java 8 function is paired with a corresponding and non-identical Java 11 function. 
    \item The Java 8 and Java 11 functions are longer than 10 lines
    \item The Java 8 and Java 11 functions are real world examples 
\end{itemize}

We conducted a deep analysis of the raw dataset. Despite initially meeting requirements such as a minimum length of 10 lines and inclusion of real-world examples, several major issues emerged during evaluation. First, the distribution of deprecated terms was highly unbalanced: most were concentrated in JAX-WS (Service) and CORBA (Any), while other deprecated APIs were largely missing. This caused the dataset to be biased toward a narrow subset of migration scenarios, limiting its generality in evaluating LLM performance.
Second, the keyword-based collection process led to a large number of false positives. Many keywords such as \texttt{Service} and \texttt{Any} frequently appeared in regular function names and variables rather than deprecated API calls. Similarly, ORB was often part of unrelated words like forbidden, and other collected terms like CommandMap and BindingProvider referred to internal project functions instead of actual deprecated APIs. These factors resulted in a large portion of invalid samples. Moreover, after expanding the scraping scope and refining keyword constraints through a more specific keyword search and manual analysis, it remained difficult to obtain reliable data, as most Java 8 repositories on GitHub had already removed deprecated APIs prior to Java 11 migrations. Consequently, the dataset failed to represent realistic or diverse migration cases, making it unsuitable for robust evaluation.

To address these limitations, a “revised” dataset was manually curated with assistance from ChatGPT for source code generation. The process began with modifying the keyword list to include a broader range of deprecated APIs while removing ambiguous terms that previously caused false positives. Using this updated list, we selected and refined 45 function pairs, each representing a clear migration from Java 8 to Java 11. These were designed to preserve functional equivalence while replacing deprecated methods (e.g., converting Collections.singletonList to List.of) or outdated APIs with their modern counterparts. These pairs contained the same number of parameters to ensure that the focus was on Java syntax changes. 

Compared to the original dataset, the revised dataset offered a more balanced distribution across eight API categories and contained significantly fewer outliers in function length. Statistical analysis (visible in Table~\ref{tab:dataset_2_stats}) showed shorter and more uniform function sizes, indicating lower noise and improved focus on deprecated API replacement. Although the dataset was synthetic and thus not directly reflective of real-world codebases, it provided a cleaner and more controlled benchmark for assessing LLMs’ ability to detect, update, and preserve functionality during Java version migration.

\begin{table}
\centering
\caption{JMig Bench Statistics}
\vspace{-0.5em}
\begin{tabular}{|l|c|c|l|}
\hline
\textbf{Metric} & \textbf{Java 8} & \textbf{Java 11} \\
\hline
Number of Functions & \multicolumn{2}{c|}{45} \\
\hline
Average Function Length & 9.69 & 8.33 \\
Maximum Function Length & 20 & 22 \\
Minimum Function Length & 4 & 3 \\
\hline
Average Number of Parameters & \multicolumn{2}{c|}{0.58} \\
\hline
\end{tabular}
\label{tab:dataset_2_stats}
\end{table}

\begin{figure}
    \centering
    \includesvg[width=\columnwidth]{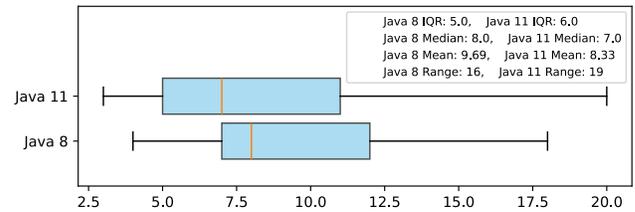}
    \caption{Function Length Boxplot (Final Dataset)}
    \label{fig:SecondaryBoxPlot}
\end{figure}

\begin{figure}
    \centering
    \includesvg[width=\columnwidth]{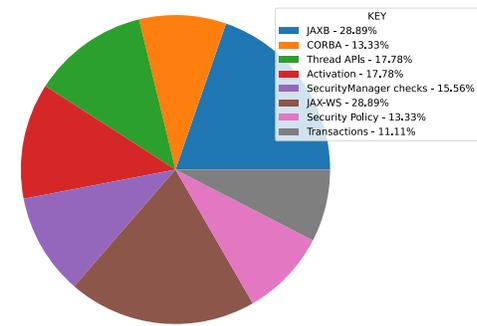}
    \caption{Distribution of Deprecated Terms (Final Dataset)}
    \label{fig:SecondaryPie}
\end{figure}

The new dataset was designed to address the imbalance and noise issues identified in the initial automatically constructed dataset. It covers a wider range of deprecated APIs compared to the web-scraped dataset, which were then grouped into eight distinct categories for clear analysis. As shown in Figure \ref{fig:SecondaryPie}, the distribution of deprecated keywords across these API categories is far more balanced, ensuring that no single library dominates the dataset. Compared to the initial version, the functions in this revised dataset are generally shorter and exhibit fewer extreme outliers in length (Figure \ref{fig:SecondaryBoxPlot}). This indicates that the new dataset contains less noise and better isolates specific deprecated API usages, providing a cleaner and more controlled benchmark for evaluating LLMs’ ability to handle Java version migration.

\begin{figure*}[!t]
\lstset{language=Python, basicstyle=\ttfamily\small, breaklines=true, frame=single, backgroundcolor=\color{gray!5}, keywordstyle=\color{blue}, stringstyle=\color{teal!70!black}, showstringspaces=false, numbers=none}
\begin{lstlisting}
prompt_messages = [
    {"role": "system", "content":
        "You are a senior Java engineer. Convert Java 8 code to Java 11 while preserving behavior."
        "Change any syntax in the Java 8 code that was removed or deprecated by Java 11."
        "Output ONLY the Java 11 method."
     },
    {"role": "user", "content":
        "Migrate the following Java 8 method (encapsulated within the <Java> and </Java> tags) to Java 11.\n\n"
        "<Java>\n" + string + "</Java>"
     },
]
\end{lstlisting}
\caption{Prompt messages used for all experiments}
\label{fig:prompt}
\end{figure*}

\section{Experimental Methodology}\label{sec:exp_method}
\subsection{Research Questions}
We want to know how good the LLM-generated migration code is and how effective it is in helping people perform migration tasks. Therefore, we propose the following RQs.

\textbf{RQ1 (Lexical and Semantic similarity):} How similar is the true Java 11 function to the generated Java 11 function?

\textbf{RQ2 (Migration Correctness):} Has deprecated syntax been removed from the Java 8 function?

\subsection{Methodology of RQ1}

The first research question focuses on the lexical and semantic similarity between LLM-generated and human-written Java 11 functions. Lexical similarity measures textual overlaps in keywords and tokens, while semantic similarity evaluates the structural and data-flow equivalence using syntax trees. Together, these metrics quantitatively assess how well the original code’s structure is preserved during migration.

Using the Python CodeBLEU package, we will calculate the \enquote{weighted ngram match score} and  \enquote{dataflow match score} values to compare the lexical and semantic similarities respectively.

\subsection{Methodology of RQ2}
This research question observes whether migration was actually performed. This consists of ensuring that deprecated syntax was correctly removed from the original Java 8 code by the LLM. While the logic preservation isn't measured, this research question uses a lightweight measurement to quantify the success of the LLM.
\\
Our calculation will use a pre-defined array of deprecated terms to calculate an inverted percentage. Dividing the number of keywords left in the generated code with the total number of keywords in the original Java 8 code.
\[
\text{Effectiveness}(\%) \;=\; 1 - \frac{\text{No. keywords in LLM generated Java 11 code}}{\text{No. keywords in original Java 8 code}}
\]

\subsection{Experimental Settings}
\subsubsection*{Codestral by Mistral AI}
We chose Mistral Codestral as the primary LLM for our experiments as it is a specialised LLM for code. It was trained on transformation and generation tasks, making it a well suited LLM for migrating Java 8 functions to Java 11 functions. Codestral also supports over 80 programming languages, making it a viable model for a wide range of developers to use in practice. Using the Mistral API expedited the experimentation stage as a local model did not need to be downloaded or instantiated. Through the API, we selected the \enquote{codestral-latest} model. At the time of experimentation (September 2025), this was Codestral 2501. 

\subsubsection*{The Prompt}
We used a standardised role-based prompt in all of our experiments. The \enquote{system} message instructed the model to act as a senior developer with the task of converting Java 8 code to Java 11 code. The \enquote{user} prompt, then clarified the conversion task and provided the LLM with the Java 8 function itself. This prompt can be seen in figure \ref{fig:prompt}. We encapsulated the Java 8 function in  \enquote{<Java>} and \enquote{</Java>} elements to create a clear structural boundary between the prompt and the code. This was aimed to prevent the model from interpreting the natural language prompt as part of the Java 8 source code. It also aimed to improve the output consistency and reduce erroneous text from making its way into the response.
\\
We intentionally evaluate a single-shot prompt, where the LLM is prompted once per function without ant iterative feedback or refinement. This design choice allows us to assess the models native ability to perform Java version migration, with no external assistance. By restricting the interaction to a single prompt, we establish a clear and reproducible baseline for automated migration.

\subsubsection*{The Dataset}
We decided to perform our experimentation on the revised dataset comprised of 45 function pairs. We made this decision as the larger dataset that was built using web scraping methods contained too much noise to yield meaningful results. The dataset does not include test cases, and therefore the generated Java code was not compiled or executed as part of our evaluation. This decision was made due to the functional-level scope of our benchmark, where individual methods are evaluated in isolation. These individual method often lack surrounding project dependencies and any build configuration files required for compilation. As a result, we focus on a static analysis based comparison metrics, deferring compilation and execution based metrics to future work.

\section{Experimental Results}\label{sec:results}
In this section, we report and interpret the results of our experiments from using the Mistral Codestral model on the JMigBench dataset to assess its effectiveness when performing Java 8 to Java 11 migrations.

\subsection{RQ1: Lexical and Semantic Similarity}

\subsubsection{CodeBLEU Metrics}

We calculated the CodeBLEU score and its sub-metrics (N-gram Match, Weighted N-gram Match, Syntax Match, and the Dataflow Match scores) to evaluate the extent to which the lexical and structural aspects of the code were preserved. Table \ref{tab:codebleu_scores} shows the scores for 40 of the 45 functions in the dataset. 5 functions had no dataflow match scores and were omitted to prevent skewing the scores (as recommended by codebleu). This happens as codebleu is unable to find any meaningful dataflows when a function is too simple and does not mean that a function was migrated incorrectly.  
\\
The table not only shows a comparison between the Generated Java 11 code and the Ground Truth Java 11 code, but also a comparison between the Java 8 code and the Ground Truth Java 11 code. We did this to evaluate whether the generated Java 11 code was closer or further to the true Java 11 code than the input Java 8 code. This comparison shows a baseline and makes it easier to interpret the scores. 

\begin{table}
\centering
\caption{Averaged CodeBLEU Metrics for 40 of 45 functions}
\label{tab:codebleu_scores}
\begin{tabular}{|l|c|c|c|}
\hline
\textbf{Metric} & 
\makecell{\textbf{  Generated vs } \\\textbf{  True Java 11}} & \makecell{\textbf{Input Java 8} \\\textbf{vs True Java 11}} \\
\hline
CodeBLEU & 0.63 & 0.62 \\
N-gram Match & 0.57 & 0.55 \\
Weighted N-gram Match  & 0.66 & 0.67 \\
Syntax Match  & 0.68 & 0.66 \\
Dataflow Match  & 0.62 & 0.59 \\
\hline
Complete Matches & \multicolumn{2}{c|}{5} \\
Omitted Functions & \multicolumn{2}{c|}{5} \\
\hline
\end{tabular}
\end{table}

Table \ref{tab:codebleu_scores} indicates that the generated Java 11 functions moved closer to the ground truth Java 11 functions than the original Java 8 function. This can be seen through the increase in scores in 3 of the 4 sub-metrics. The CodeBLEU score itself went from 0.62 (for the Java 8 code) to 0.63 (for the generated code). 
\\
Lexical similarity also increased. This can be seen in the N-gram match score which increased from 0.55 to 0.57 showing improved alignment with lexical and textual similarity. The table shows that the weighted N-gram match score dropped by 0.01, however, this was due to rounding as the raw figures were 0.664 and 0.667 respectively. The numbers were effectively the same which confirms that lexical similarity was maintained during source code migration. 
\\
Semantic similarity can be measured using the syntax match score (which rose from 0.66 to 0.68) which shows that the migrated functions had better structural similarity to the true Java 11 code than the initial Java 8 function. The dataflow match score also confirms this as it increased from 0.59 to 0.62, proving that the execution paths in the generated code was also preserved and improved in comparison to the Java 8 code. 
\\
Five functions (11.11\%) were identified as a complete match. In this case, the generated Java 11 code was identical (lexically and semantically) to the true Java 11 code. These were identified as their codeblue score and all the sub-metrics had a score of 1. Upon manual examination, the 5 functions were trivial as they only required simple one to one API replacements with no major structural changes. It shows potential that the Codestral was able to perform perfectly correct migrations, however, as only 11.11\% of the functions were migrated perfectly, the consistency of the Codestral and its ability to work on complex functions could be questioned.

\subsubsection{Example Output Function}
One example of a Java 11 function that was generated by Codestral can be seen in figure \ref{fig:diffjava11gen}. The LLM successfully identified the change on line 10 which was also made in the true Java 11 function. \enquote{Collections.singletonList} was changed to \enquote{List.of}. Codestral also made a similar change on line 8 of the function where \enquote{Arrays.asList} became \enquote{List.of}. This change would make the \enquote{isp} list immutable when it was previously a mutable list. In this instance the program would still function as intended as \enquote{isp} was not modified later on in the function. The LLM also removed an empty line on line 13, a subtle change with no impact on the functionality of the code.

\begin{mdframed}
\textbf{Answer for RQ1:} The findings demonstrate that, on average, the generated code had a moderate alignment to the ground truth. This was reinforced by the fact that the scores increased when comparing the Java 8 code and the generated Java 11 code to the ground truth. It shows that Codestral was able to make improvements to the Java 8 code in the right direction. The fact that only 11.11\% of the functions (which were trivial) achieved a perfect match score indicates that this model exhibits limited consistency on more complex tasks within the JMigBench dataset.

\end{mdframed}

\subsection{RQ2: Migration Correctness}
\begin{figure}
    \centering
    \includesvg[width=\columnwidth]{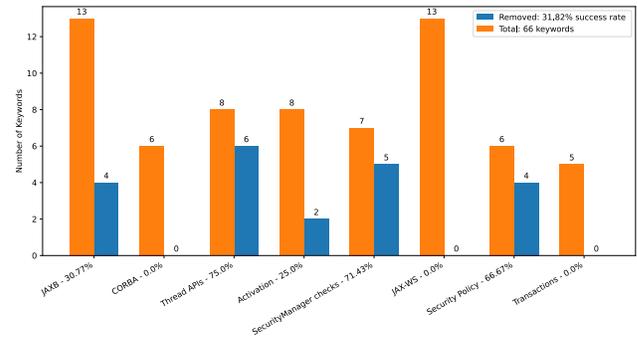}
    \caption{Bar Chart showing the keyword removal success}
    \label{fig:KeywordBars}
\end{figure}

This research question examines whether the keywords that were identified as deprecated from Java 11 were actually removed from the Java 8 code. This is important as if a deprecated keyword is still present in the generated Java 11 code, the code would fail to compile and would be principally incorrect. 
\\
Figure \ref{fig:KeywordBars} presents the distribution of deprecated keywords across the secondary dataset. Each bar represents a category of Java 8 APIs that were deprecated in Java 11. We will use this bar chart to examine the overall success of keyword removal and identify areas where Codestral excelled and where it fell short. 

At a granular level, the removal rate of deprecated keywords was 31.8\% showing that Codestral was able to remove some of the deprecated terms, but frequently left others untouched. The level of imbalance is shown in Figure \ref{fig:KeywordBars} which shows a breakdown of keyword removal across 8 categories. 
\\
Categories such as Thread APIs (75.0\%), SecurityManager checks (71.43\%) and Security Policy related keywords (66.67\%) were consistently migrated to a relatively high degree of success. In contrast, CORBA, JAX-WS and Transaction related keywords always remained in the generated code. The bar chart shows that there were no examples in which these APIs were removed from the Java 8 code.

Upon manual inspection, we noticed the model was able to identify simple cases where trivial substitutions could be performed, however, where more complex reasoning was required, the model failed to make any changes. We noticed that in these cases, the failed data items had identical CodeBLEU scores for the Java 8 and the generated Java 11 code (when being compared to the ground truth). This showed that Codestral failed to apply any migration changes in these cases under the evaluated experimental settings.

\begin{mdframed}
\textbf{Answer for RQ2:} Our experiment shows that Codestral achieved a limited level of success with keyword removal. 40\% of the functions had all keywords removed, however, only 31.82\% of the 66 keywords used in the dataset were successfully removed. 
An area of limitation was highlighted in the fact that certain deprecated APIs were harder to migrate than others. 
\end{mdframed}

\balance

\subsection{Discussion}

Our findings from RQ1 and RQ2 provide compatible perspectives on the effectiveness of automated migration from Java 8 to Java 11 using Codestral on the JMigBench benchmark. RQ1 showed that the generated code achieved moderate similarity to the true Java 11 and modified the code in the right direction. It also showed 5 cases where migration was performed perfectly. RQ2 however highlighted some inconsistencies within the results which showed that only 40\% of the functions were migrated correctly. When analysed through a stricter, more categorial view, it showed that migration was only successful in simpler and more trivial cases, and the LLM struggled to work when more complex changes were required (e.g. with CORBA). 
\\
Our finding is that there is an uneven level of success when using Codestral to perform this type of source code migration within the scope of our benchmark. On one hand, Codestral could reduce human effort with trivial cases by automating common migrations; however, due to its inconsistencies on more complex transformations, human intervention remains essential for this model and our benchmark configuration.
\\

\section{Threats to Validity}\label{sec:threats}

\paragraph{Threats to Internal Validity.} A key limitation of this study is the absence of test execution or runtime validation. The benchmark operates at a functional level, where individual methods are evaluated in isolation. The lack of surrounding project context, external dependencies and build configurations required for compilation and testing means that runnable test cases were not viable for the functions in the dataset. 

Consequently compilation and execution based validation were out of scope for this work, and our evaluation relied on static analysis metrics such a CodeBLEU and keyword-based migration correctness. While these metrics provide useful insights into lexical, syntactic, and structural changes introduced during migration, they cannot guarantee compilation success or functional equivalence. This limitation is inherent to function-level migration benchmarks and highlights the trade-off between fine-grained analysis and executable completeness.

\paragraph{Threats to External Validity.} Our experiments were solely performed on the Mistral Codestral LLM. This model is specialised to code generation and, therefore, well-suited to the task of source code migration. However, the use of a single LLM limits the findings to Codestral. %

\section{Conclusion}\label{sec:conclusion}
This research project aims to evaluate the effectiveness of LLMs for migrating Java 8 code to Java 11 code, while building a benchmark to support systematic evaluation in the future. Using this dataset, our evaluation showed that Codestral performed well on more trivial functions were one-to-one substitutions could be made, however, failed to function when more complex APIs were introduced. This alerted us to the fact that the Codestral demonstrated limited reasoning when it had to implement functions involving more complex APIs or multi-step transformations. Our key finding is that {Codestral can be useful for simple migrations, however, further research will be needed to build an LLM which can handle complex migrations with a greater level of consistency. In our experiments, Codestral achieved perfect migrations in only 11.11\% of cases, leaving room for improvement to achieve perfect migrations in 90\%-95\% of cases.

\appendix

\bibliographystyle{ACM-Reference-Format}
\bibliography{JMigBench_Bib}

\end{document}